\documentclass[conference]{IEEEtran}
\IEEEoverridecommandlockouts
\usepackage{cite}
\usepackage{amsmath,amssymb,amsfonts}
\usepackage{algorithm, algorithmic}
\usepackage{algorithmic}
\usepackage{graphicx}
\usepackage{bm}
\usepackage{textcomp}
\usepackage{xcolor}
\def\BibTeX{{\rm B\kern-.05em{\sc i\kern-.025em b}\kern-.08em
    T\kern-.1667em\lower.7ex\hbox{E}\kern-.125emX}}
\begin{document}

\title{Pilot Allocation for Multi-Hop Over-the-Air Neural Inference under Imperfect CSI}

\author{\IEEEauthorblockN{Tolga Girici}
\IEEEauthorblockA{\textit{Dept. of Electrical and Electronics Engineering} \\
\textit{TOBB University of Economics and Technology}\\
Ankara, Türkiye \\
tgirici@etu.edu.tr}
\and
\IEEEauthorblockN{Meng Hua and Deniz Gündüz}
\IEEEauthorblockA{\textit{Dept. of Electrical and Electronics Engineering} \\
\textit{Imperial College London}\\
\textit{United Kingdom} \\
\{m.hua, d.gunduz\}@imperial.ac.uk
}}

\maketitle

\begin{abstract}
A multi-hop amplify-and-forward (AF) relay network can emulate a fully connected (FC) neural network layer via over-the-air (OTA) computation. However, achieving high emulation accuracy requires accurate channel state information (CSI) across all links in the multi-hop network. In this work, we investigate the impact of CSI errors on classification performance. We propose five heuristic schemes for allocating the total channel training time (pilots) across hops and compare their effectiveness. Numerical results reveal a clear trade-off between channel training overhead and classification accuracy. In particular, with sufficient pilot power and balanced allocation of channel training resources, the system  can achieve classification accuracy close to that of the digital baseline.
\end{abstract}


\section{Introduction}

Modern edge applications, including augmented reality, autonomous systems, and massive IoT, require low-latency inference under strict energy and communication constraints. Offloading inference to the cloud introduces significant latency and communication overhead due to the transmission of raw sensor data and intermediate features. Over-the-air (OTA) computation offers a promising alternative by exploiting the superposition property of wireless channels to directly compute functions in the air, thereby avoiding quantization, packetization, and reconstruction overheads. The foundations of OTA computation are well established in the literature \cite{goldenbaum2013robust}, and a comprehensive survey is provided in \cite{csahin2023survey}.

Building on this paradigm, OTA computation has been increasingly applied to machine learning tasks \cite{hua2026wireless}. In particular, OTA-based aggregation has been used for distributed and federated learning \cite{amiri2020federated}, enabling efficient model updates over wireless channels. More recently, OTA inference and split learning frameworks have been proposed for wireless MIMO systems \cite{yang2023over}, while analog OTA machine learning techniques have further strengthened the integration of communication and learning \cite{yilmaz2025private},\cite{zhu2024over}. These works demonstrate that OTA computation can significantly reduce communication overhead and latency in learning-based applications, in both training and inference stages.

Beyond data aggregation, OTA computation has also been used to directly implement neural network operations. The AirFC framework \cite{reus2023airfc} demonstrates how a fully connected (FC) layer can be realized over the air using wireless signals. RIS-assisted approaches have also been proposed to implement neural network layers, including AirNN \cite{sanchez2022airnn} and other RIS-based neural architectures \cite{hua2026implementing}, where the effective wireless channel is engineered to mimic the weight matrix of a neural network. These approaches highlight the potential of analog wireless computation for neural inference, but also reveal challenges related to channel rank, noise accumulation, and hardware complexity. In particular, RIS-based implementations require careful deployment and may incur significant cost when multiple surfaces are needed to achieve full-rank transformations.

Amplify-and-forward (AF) relaying has also been studied in the context of OTA computation to improve signal quality and extend coverage. Prior works have investigated relay-assisted OTA computation in IoT networks \cite{wan2023energy}, scheduling strategies for AF-based OTA systems \cite{tang2022node}, and hierarchical aggregation via relays \cite{wang2022amplify}. However, these works focus primarily on data aggregation and do not address the implementation of neural network layers. More recently, multi-hop OTA inference over MIMO systems has been considered in \cite{bian2025over}, demonstrating the feasibility of performing inference over cascaded wireless channels. In our recent work \cite{girici2026realization}, we studied optimal power allocation and precoding in OTA neural network implementation in a multi-hop AF relaying network.  

Another important limitation in the existing literature is the common assumption of perfect channel state information (CSI). In practice, CSI must be acquired through pilot-based estimation and is inherently imperfect. Recent studies have shown that CSI errors can significantly degrade OTA computation performance. For example, \cite{chen2022over} demonstrates that imperfect CSI leads to an irreducible mean-square error even at high transmit power, and proposes robust power control and beamforming strategies. This is extended to multi-carrier systems in \cite{chen2023over}, where joint optimization under CSI uncertainty is considered. Despite these advances, the impact of CSI acquisition and pilot resource allocation in multi-hop OTA neural inference has not been thoroughly investigated.

While prior work on OTA neural inference typically assumes either perfect CSI or single-hop channels, the problem of joint CSI acquisition and channel training time (i.e., pilot) allocation across multi-hop architectures remains largely open. In multi-hop systems, channel estimation errors accumulate across hops, and the allocation of limited channel training resources across layers becomes a critical design consideration.

In this paper, we address these challenges by studying the implementation of a FC neural network layer over a multi-hop AF relay network under imperfect CSI. We propose a pilot-based channel estimation framework and investigate how the total  channel training time should be allocated across hops. We introduce five heuristic training time allocation strategies and evaluate their impact on the accuracy of OTA neural inference. Our results reveal important trade-offs among channel training overhead, CSI quality, and inference accuracy, providing new insights for the design of multi-hop OTA neural computing systems.

\section{System Model}

\begin{figure}[htbp]
\centerline{\includegraphics[width=\columnwidth]{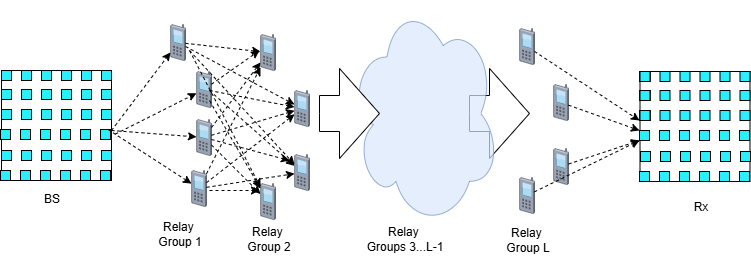}}
\caption{Multi-hop OTA computing system model \cite{girici2026realization}}
\label{fig:MultiHopAirFC}
\end{figure}

We assume a multi-antenna base station (BS) with $N_t$ antennas transmitting to a multi-antenna receiver (Rx) through $K$ single-antenna AF relay devices, as shown in Fig. \ref{fig:MultiHopAirFC}. Relay devices are randomly distributed over an area,  organized geographically into $L$ relay groups in series, where group $l$
consists of $K_l$ single-antenna AF relays (i.e. $\sum_{l=1}^L K_l=K$). Let $\mathbf{H}_{1} \in \mathbb{C}^{K_1\times N_t}$ denote the complex  baseband channel matrix from the BS to the first relay group, where $[\mathbf{H}_{1}]_{i,j}=h_{i,j}^1$ is the channel gain from the $j^{th}$ antenna of the BS to the $i^{th}$ device in group $1$.

Let $\mathbf{x}\in \mathbb{C}^{N}$ be the transmitted baseband complex signal vector. The BS precodes $\mathbf{x}$ using precoding matrix $\mathbf{F}_1\in\mathbb{C}^{N_t\times N}$.  Suppose that, upon receiving the precoded signal from the BS, each device in relay group $1$ amplifies and forwards this signal to the second relay group. Upon receiving the signal, relay device $k$ in group $l$ amplifies the signal with complex weight $a_k$ and forwards it to the next stage.  Let us define the diagonal forwarding matrix of relay group $l$ as  $\mathbf A_l \triangleq \mathrm{diag}(\mathbf a_l)\in\mathbb C^{K_l\times K_l}$ with
$\mathbf a_l\in\mathbb C^{K_l}$.  We assume that there is perfect synchronization among the AF relay devices in a group.

We assume that the BS and each relay group share the wireless channel in a time-division multiple access (TDMA) manner.   Let $\mathbf H_{l+1} \in \mathbb C^{K_{l+1}\times K_l}, l=1,\dots,L-1$  denote the channel matrix between relay groups $l$ and $l+1$. Here $[\mathbf{H}_{l}]_{i,j}=h_{i,j}^l$ is the channel gain from device $i$ in group $l$ to device $j$ in group $l+1$.  Lastly, the $L$-th  relay group transmits to the Rx, which has $N_r$ receive antennas.  Let the column vector $\mathbf{g}_k  \in \mathbb{C}^{N_r\times 1}$ be the complex baseband channel from device $k$ in group $L$ to the Rx, which are collected into $\mathbf H_{L+1}=[\mathbf{g}_1,\ldots,\mathbf{g}_K]\in\mathbb{C}^{N_r\times K_L}$. 

A direct channel from the BS to the Rx may also exist, with channel matrix $\mathbf{H}_{0} \in \mathbb{C}^{N_r\times N_t}$. Then, the effective baseband channel of the system can be defined as follows:
\begin{equation}
  \mathbf H_{\rm eff}
  = \mathbf H_{0} +\mathbf H_{L+1}\mathbf A_L \mathbf H_L \cdots \mathbf A_2 \mathbf H_2 \,\mathbf A_1 \mathbf H_1
  \ \in \ \mathbb C^{N_r\times N_t}.
\end{equation}

Finally, the received signal at Rx is multiplied by a complex combining matrix  $\mathbf{F}_2 \in \mathbb{C}^{N_r\times N }$. The received signal at the Rx becomes,
\begin{equation}
  \mathbf y \ =\ \mathbf F_2\big( \mathbf H_{\rm eff}\,\mathbf F_1 \mathbf x + \mathbf n_{\rm in}\big),
  \label{eq:inputoutput}
\end{equation}
where $\mathbf n_{\rm in}$ is the cumulative noise at the Rx input, which is derived as follows: Each relay group $l$ introduces  noise $\mathbf n_l\sim\mathcal{CN}(\mathbf 0,\sigma_{u,l}^2\mathbf I_{K_l})$
\emph{before} amplification, while $\mathbf n_c\sim\mathcal{CN}(\mathbf 0,\sigma_c^2\mathbf I_{N_r})$ is added at the Rx. Due to linear operations $\mathbf n_{\rm in}$ is also Gaussian with $\mathcal{CN}(\mathbf 0,\mathbf R_n^{\rm in})$, where the aggregate noise covariance at the Rx is, 
\begin{equation}
\label{eq:Rn_in}
  \mathbf R_n^{\rm in}
  \ =\ \sigma_c^2\,\mathbf I_N \ + \ \sum_{j=1}^{L}\ \mathbf T_j\,\sigma_{u,l}^2\,\mathbf T_j^{H}.
\end{equation} Here $\mathbf T_j$ defines the transfer matrix from group $j$ to the Rx input:
\begin{multline}
  \mathbf T_j \ \triangleq\ \mathbf H_{L+1}\mathbf A_L \mathbf H_L \cdots \mathbf H_{j+1}\mathbf A_j
  \ \in\ \mathbb C^{N_r\times K_j},
  \qquad \\j=1,\dots,L.
\end{multline} These expressions show how  noise  accumulates at the Rx due to multi-hop AF relaying.

\subsection{Channel Estimation}

We assume pilot-based channel estimation, where transmitters send pilot signals to acquire CSI.

Let $\tau_p^l$ be the pilot length (number of channel uses) allocated to the devices in group $l$, where $\tau_p^0$ and $\tau_p^L$ denote the pilot lengths used by the BS and the group $L$ devices, respectively. Since we assume orthogonal pilots, the pilot length must be at least as large as the number of antennas at each layer. Due to TDMA-based channel access, the same pilot sequences can be reused across the BS and each group of relays. Hence, the minimum required pilot dictionary size becomes $\tau_p^{\mathrm{dict}}=\max\{N_t,K_1, K_2,\ldots,K_L \}$. The minimum required training time for each hop is as follows, 
\begin{align}
    &\tau_p^0\geq N_t, \nonumber\\
    &\tau_p^l\geq  K_l, l=1,...,L. 
\end{align}

\subsubsection{Estimation of inter-group channels}

Device $k$ in group $l$ sends the pilot sequence $\boldsymbol{\phi}_{k}^l\in \mathbb{C}^{\tau_{p}^l\times 1}$, where $||\boldsymbol{\phi}_{k}^l||^2=\tau_{p}^l$. We assume a constant pilot transmit power $p_p$ over the whole network \footnote{Optimizing the pilot power across hops is a possible direction for future research.}. Let $\mathbf{h}^{l}_m\in \mathbb{C}^{1\times K_l}$ be the channel from relay devices in group $l$ to device $m$ in group $l+1$.  Then the received pilot vector at device $m$ in group $l+1$ becomes,
\begin{equation}
\mathbf{y}_{m}^{l+1}=\sum_{k=1}^{K_l}\sqrt{p_p}h^{l}_{m,k}\mathbf{\phi}_{k}^l+\mathbf{n}_{l+1},
\end{equation}where $\mathbf{n}_{l+1}\sim\mathcal{CN}(\mathbf{0},\sigma_{u,l}^2 \mathbf{I}_{\tau_{p}^l})$. After receiving this signal, device $m$ in group $l+1$ correlates this signal with $\mathbf{\phi}_{k}^l$. The least-squares (LS) estimate is given by,
\begin{equation}
    \hat{h}^{l}_{m,k}=\frac{1}{\sqrt{p_p}\tau_p^l}\mathbf{\phi}_{k,l}^H\mathbf{y}_{m}^{l+1}.
\end{equation}By correlating one-by-one with $\mathbf{\phi}_{1}^l, \mathbf{\phi}_{2}^l, \ldots, \mathbf{\phi}_{K_l}^l$, device $m$ in group $l+1$ estimates the $m^{th}$ row of $\mathbf{H}_{l+1}$ as $\hat{\mathbf{h}}^{l}_{m}=[\hat{h}^{l}_{m,1}, \hat{h}^{l}_{m,2}, \ldots, \hat{h}^{l}_{m,K_l}]$. These estimates are sent by individual devices  to a center and the rows are combined to get the estimate $\hat{\mathbf{H}}_l$ of the channel matrix between groups $l $ and $l+1$.

Channel estimations for the BS-group 1 link and the group L - Rx link are done similarly. 

\subsubsection{Estimation of the BS-to-group 1 channel}

The BS employs orthogonal pilot sequences across its $N_t$ antennas. Let the pilot matrix transmitted by the BS be
\begin{equation}
\mathbf{\Phi}^{0}=
\begin{bmatrix}
\boldsymbol{\phi}^{0}_{1}, & \boldsymbol{\phi}^{0}_{2}, & \ldots ,& \boldsymbol{\phi}^{0}_{N_t}
\end{bmatrix}
\in\mathbb{C}^{\tau_p^0\times N_t},
\end{equation}
with $(\mathbf{\Phi}^{0})^H\mathbf{\Phi}^{0}=\tau_p^0 \mathbf{I}_{N_t}$. Then the received pilot vector at relay device $m$ in group 1 is
\begin{equation}
\mathbf{y}^{1}_{m}=\sqrt{p_p}\,\mathbf{\Phi}^{0}\mathbf{h}^{0\,T}_{m}+\mathbf{n}^{1}_{m},
\end{equation}
where $\mathbf{h}^{0}_{m}\in\mathbb{C}^{1\times N_t}$ is the channel row from the BS antennas to relay $m$ in group 1, and $\mathbf{n}^{1}_{m}\sim\mathcal{CN}(\mathbf{0},\sigma_{u,1}^2\mathbf{I}_{\tau_p^0})$. By correlating $\mathbf{y}_m^1$ with the BS pilots, relay $m$ obtains the LS estimate
\begin{equation}
\hat{\mathbf{h}}^{0}_{m}
=
\frac{1}{\sqrt{p_p}\tau_p^0}
(\mathbf{\Phi}^{0})^H \mathbf{y}^{1}_{m}.
\end{equation}
Stacking these estimated rows for all $m=1,\ldots,K_1$ yields the estimate $\hat{\mathbf{H}}_{1}\in\mathbb{C}^{K_1\times N_t}$ of the BS-to-group 1 channel matrix $\mathbf{H}_1$.

\subsubsection{Estimation of the group $L$-to-Rx channel}

To estimate the channel from the last relay group to the Rx, the relays in group $L$ transmit orthogonal pilot sequences. Let
\begin{equation}
\mathbf{\Phi}^{L}=
\begin{bmatrix}
\boldsymbol{\phi}^{L}_{1} & \bm{\phi}^{L}_{2} & \cdots & \boldsymbol{\phi}^{L}_{K_L}
\end{bmatrix}
\in\mathbb{C}^{\tau_p^{L}\times K_L},
\end{equation}
with $(\mathbf{\Phi}^{L})^H\mathbf{\Phi}^{L}=\tau_p^{L}\mathbf{I}_{K_L}.$
The received pilot signal at the Rx is then
\begin{equation}
\mathbf{Y}_{\rm Rx}
=
\sqrt{p_p}\,\mathbf{G}\,(\mathbf{\Phi}^{L})^{T}
+
\mathbf{N}_{\rm Rx},
\end{equation}
where $\mathbf{Y}_{\rm Rx}\in\mathbb{C}^{N\times \tau_p^{L}}$ and $\mathbf{N}_{\rm Rx}\sim\mathcal{CN}(\mathbf{0},\sigma_c^2\mathbf{I}_{N}\otimes \mathbf{I}_{\tau_p^{L}}).
$
The LS estimate of the channel matrix $\mathbf{G}=\mathbf{H}_{L+1}$ is
\begin{equation}
\hat{\mathbf{G}}
=
\frac{1}{\sqrt{p_p}\tau_p^{L}}
\mathbf{Y}_{\rm Rx}
(\mathbf{\Phi}^{L})^{*}.
\end{equation} The obtained CSI is transmitted to a central node (e.g. the BS) in a multi-hop manner and used for the optimal implementation of the OTA FC layer. Channel estimation is performed once before each inference event (i.e., a forward pass of the NN), and the channel is assumed to remain fixed during inference. 

\section{Multi-Hop AirFC with AF}
Let $\mathbf W\in\mathbb C^{N_r\times N_t}$ be the target weight matrix of the FC layer. Our goal is to design the OTA system so that the input-output relation in  (\ref{eq:inputoutput}) approximates the FC layer, i.e.,  
\begin{equation}
  \label{eq:fc}
  \mathbf{y} = \mathbf{W}\mathbf{x} + \mathbf{b},\qquad
  \mathbf{x},\mathbf{y}\in\mathbb{C}^{N_t\times 1},\; \mathbf{W}\in\mathbb{C}^{N_r\times N_t}.
\end{equation}A BS with $N_t$ antennas precodes $\mathbf{x}$ using precoding matrix $\mathbf{F}_1\in\mathbb{C}^{N_t\times N_t}$  and the relay devices in group $l$ amplify-and-forward their received signals using gain matrix $\mathbf{A}_l$. Then, the received signal from the $L$-th relay group is multiplied by a complex combining matrix  $\mathbf{F}_2 \in \mathbb{C}^{N_r\times N }$ at the receiver. The output signal becomes,

\begin{equation}
  \label{eq:r}
  \mathbf{y} = \mathbf{F}_2\hat{\mathbf{H}}_{\rm eff}\mathbf{F}_1\mathbf{x} + \mathbf{F}_2\mathbf{n}_{in},
\end{equation} where $\hat{\mathbf{H}}_{\rm eff}$ is the estimated effective baseband channel of the system, involving the estimated BS-group 1, group-L-Rx and inter-group channel matrices. We note that the bias term $\mathbf{b}$ in (\ref{eq:fc}) is not implemented over-the-air.  In this work, we focus on the OTA realization of the linear transformation $\mathbf{Wx}$, and assume that the bias term is applied digitally at the Rx after OTA computation.

We previously formulated and solved the multi-hop AirFC problem under perfect CSI in \cite{girici2026realization}; the formulation is briefly recapped below for the sake of completeness. 

\subsection{Imitation Objective and Constraints}

We seek the OTA parameters $(\mathbf F_1,\mathbf F_2,\{\mathbf A_l\})$ that minimize
an imitation error plus a noise penalty term:
\begin{subequations}
\label{eq:airfc_problem}
\begin{align}
  \min_{\mathbf F_1,\mathbf F_2,\{\mathbf A_l\}}\
  & \underbrace{\big\|\,\mathbf F_2 \hat{\mathbf{H}}_{\rm eff} \mathbf F_1 - \mathbf W\,\big\|_F^2}_{\text{FC imitation error}}
    \ +\ \underbrace{\mathrm{tr}\!\big(\mathbf F_2 \mathbf R_n^{\rm in} \mathbf F_2^H\big)}_{\text{noise propagation}}
  \label{eq:main_problem}\\
  \text{s.t.}\quad
  & \|\mathbf F_1\|_F^2 \ \le\ P_{\max}, \label{eq:tx_power}
  \\
  & \mathbb E\!\left[\,|a_{l,k} u_{l,k}|^2\,\right]\ \le\ P_{l,k}, \nonumber\\
  & \quad  \quad \quad \forall\,l=1,\dots,L,\ \ k=1,\dots,K_l. \label{eq:relay_power}
\end{align}
\end{subequations}
In \eqref{eq:relay_power}, $u_{l,k}$ denotes the (complex) signal incident on relay $(l,k)$ before
amplification. A conservative instantaneous surrogate for the variance of the incident signal is
\begin{equation}
  \label{eq:relay_input_power}
  p^{\rm in}_{l,k}\ \approx\
  \left\|\big(\mathbf H_l \mathbf A_{l-1}\mathbf H_{l-1}\cdots \mathbf A_1 \mathbf H_1 \mathbf F_1\big)_{k,:}\right\|_2^2
  \ +\ \sigma_{u,l}^2,
\end{equation}
so that $|a_{l,k}|^2\,p^{\rm in}_{l,k}\le P_{l,k}$.

We add a noise penalty to the objective because, even if the effective channel $\mathbf F_2 \mathbf H_{\rm eff} \mathbf F_1$ resembles $\mathbf W$,  noise may be amplified at the Rx, to the point that the useful signal is buried.

\subsection{Alternating Optimization (AO) Based Solution}
In \cite{girici2026realization}, we solved the optimization problem in (\ref{eq:airfc_problem}), (\ref{eq:tx_power}), (\ref{eq:relay_power}) using alternating optimization (AO), where we iteratively optimize with respect to 1) BS precoder $\mathbf{F}_1$, 2) Rx combiner $\mathbf{F}_2$, and 3) relay AF gains $\mathbf{A}_l, l=1,\ldots,L$. 

Recall the estimated  channel matrices
\(
\hat{\mathbf{H}}_1\in\mathbb C^{K_1\times N},\,
\hat{\mathbf{H}}_{l+1}\in\mathbb C^{K_{l+1}\times K_l}\ (l=1,\dots,L-1),\,
\hat{\mathbf{H}}_{L+1}\in\mathbb C^{N\times K_L}
\),
and optionally the estimated direct link \(\hat{\mathbf{H}}_0\in\mathbb C^{N\times N}\). The original problem is jointly non-convex in $\mathbf{F}_1$, $\mathbf{F}_2$, and $\mathbf{A}_l, l=1,\ldots,L$. Following \cite{girici2026realization}, we use AO, optimizing one block at a time while fixing the others.

\begin{enumerate}
\item The $\mathbf{F}_1$ subproblem is a convex quadratically constrained LS problem with a closed-form solution plus bisection for the power constraint,
\item The $\mathbf{F}_2$ subproblem is an unconstrained convex quadratic problem with a closed-form MMSE-like solution,

\item Each $\mathbf{A}_l$ subproblem ( $l=1,\ldots,L$) is a regularized LS problem, followed by projection onto per-relay power constraints.
\end{enumerate}

\section{Training Time Allocation}

Due to the TDMA-based access among the BS and relay groups, pilots transmitted at different hops do not interfere with each other. Additionally, we assume orthogonal pilot sequences so that devices in a group also do not interfere with each other. Under this assumption, the required pilot dictionary size is determined by the largest transmitting array among all training phases.  Since channel training is carried out over separate TDMA phases, the minimum total training duration equals the sum of the minimum pilot lengths required across all hops. The minimum feasible pilot lengths are
\begin{equation}
\tau_0^{\min}=N_t,\qquad \tau_l^{\min}=K_l,\; l=1,\dots,L.    
\end{equation}
Hence, the minimum total training time is
\begin{equation}
\tau_{\min}=N_t+\sum_{l=1}^{L}K_l.    
\end{equation}
As an example, for $N_t=49$, $K=120$, and $L=3$ with equal group sizes $K_1=K_2=K_3=40$, only $\max\{49,40,40,40\}=49$ orthogonal pilot sequences are required in the pilot dictionary, while the minimum total training time is $\tau_{\min}=49+40+40+40=169 $ channel uses. By contrast, when $L=1$, the same total minimum training time is obtained, namely $\tau_{\min}=49+120=169$, but in that case the pilot dictionary must support $120$ orthogonal pilot sequences.

To improve estimation accuracy, the pilots assigned to a given hop may be repeated multiple times. Let
\[
\tau_l = m_l \tau_l^{\min}, \qquad l=0,1,\dots,L,
\]
where $m_l\ge 1$ denotes the \emph{repetition factor} for hop $l$. Then the total training time becomes
\begin{equation}
\tau_{\mathrm{tot}}=\sum_{l=0}^{L}\tau_l
=\sum_{l=0}^{L} m_l \tau_l^{\min}.    
\end{equation}

Under orthogonal pilot-based LS estimation, the channel estimation error decreases approximately inversely with the effective pilot energy, $p_p\tau_l$. Therefore, allocating additional channel training time to a hop improves the CSI quality of that hop, but also increases the total training overhead. This creates a training-time allocation tradeoff across the multi-hop network. In order to evaluate the effect of training time allocation on the imitation accuracy, we considered a number of heuristics. These are multi-step greedy heuristics: at each step,  the layer $l^*$ with the highest priority is selected and its repetition factor $m_{l^*}$ is incremented by one. The heuristics start by setting $m_l=1$  for all layers. The heuristics are as follows: 

\begin{enumerate}
\item Uniform excess allocation: This heuristic distributes the excess training time as evenly as possible across hops, in terms of the repetition factor.  Each hop has weight $w^{(t)}_l=\frac{1}{m^{(t)}_l}$, at step $t$, where $m^{(t)}_l$ is the repetition factor of hop $l$ at step $t$ of the algorithm. The maximizing hop's repetition factor is incremented by one. 

\item Proportional-to-minimum allocation: 
This heuristic allocates excess pilots proportional to $\tau_l^{min}$. Each hop has weight $w^{(t)}_l=\frac{\tau_l^{min}}{m^{(t)}_l}$ at step $t$.

\item Front-loaded allocation: This heuristic allocates more excess pilots to earlier hops, motivated by error propagation. Each hop has weight $w_l=\frac{1}{(l+1)m^{(t)}_l\tau_l^{min}}, l=0,\ldots,L$ at step $t$. At each step, maximum-weight hop's repetition factor is incremented by one. By giving earlier hops more training time, we avoid poor early-hop CSI degrading all downstream design steps.

\item All-excess-to-first-hop: This heuristic assigns all excess channel training time to the first hop.

\item Channel-strength-aware heuristic: This heuristic allocates more training time to weaker hops. For example, if $\beta_l$ is an average large-scale gain or average channel Frobenius norm for hop  $l$, we set the weights as $w_l^{(t)}= \frac{1}{\beta_lm^{(t)}_l\tau_l^{min}},  l=0,\ldots,L$, since estimation becomes harder for noisy and weak hops. The scheduling metric also involves the training time allocation up to the step $t$, in order to avoid allocating all excess training time to a single hop.   We assume that the large-scale fading (i.e., pathloss) changes very slowly and is known centrally.
\end{enumerate}

\section{Numerical Results}
In this section, we numerically evaluate the impact of channel estimation errors and the channel training-time allocation heuristics on multi-hop AirFC  performance. The simulation parameters are summarized in Table~\ref{tab:simparam}. We consider a rectangular coverage area of dimensions $D_{\max}\times D_{\max}$ meters, which is divided into $L$ consecutive regions of size $D_{\max}\times \frac{D_{\max}}{L}$ along the transmission path between a multi-antenna BS and a multi-antenna Rx. Each region contains a \emph{group} of $\tfrac{K}{L}$ single-antenna AF relay devices. The wireless links between the BS, relays, and Rx follow the 3GPP UMi Street Canyon pathloss model, while the inter-relay links are modeled according to the TR 38.901 sidelink channel specifications. The line-of-sight (LoS) probability for each link is determined as a function of the link distance.

The end-to-end neural network employed in the experiments consists of an input layer, a convolutional layer with $2$ output channels (kernel size $3$, stride $4$, padding $1$), and a real-to-complex (R2C) transformation that maps real-valued inputs to complex-valued representations of reduced dimensionality. This is followed by a complex-valued FC layer with complex ReLU activation, complex batch normalization, and a power normalization layer. The signal then passes through the wireless front-end, including precoding at the BS, multi-hop AF relaying, and combining at the receiver. Finally, the processed signal passes through a complex ReLU activation, a complex-to-real (C2R) transformation, a real-valued FC layer, and the output layer.

All results are averaged over $40$ independent channel realizations. The figures include error bars to indicate the variability around the mean accuracy. The dashed black line in each plot represents the accuracy of a fully digital baseline, which serves as a reference for comparison. This baseline is obtained by training the network on the Fashion-MNIST dataset, achieving an accuracy of $84.5\%$. The goal of the  multi-hop OTA FC implementation is to approach this benchmark performance.
\begin{table}[t!]
    \centering
    \caption{Simulation Parameters}
    \begin{tabular}{|c|c|}
    \hline
        \textbf{Parameter} & \textbf{Value}  \\\hline
        Relay devices per group & $K=6,12,..., 54, 60$  \\\hline
        Number of relay groups & $L=5$  \\\hline        
        Number of antennas  &  $N=N_t=N_r=49$ \\\hline 
        BS Power  & $P_{max}=N$ W\\\hline
        Relay Power & $P_{k}=0.1, 1$ W \\\hline
        Pilot transmission power & $p_{p}=0.1, 1$ W \\\hline
        Excess training time& $200,...,1000$ channel uses\\\hline
        Carrier frequency & $f_c=28$ GHz \\\hline
        Noise PSD & $N_o=-174$dBm \\\hline
        Bandwidth & $B=300$MHz  \\\hline
        BS/Rx height & $5$ meters\\\hline
        AF relay height & $1.5$ meters\\\hline
        Network diameter & $D_{max}=100, 200$m \\\hline
        Pathloss & 3GPP UMi Street Canyon (NLoS)\\\hline
        BS-Rx MIMO Channel & Ricean ($\kappa=0,$ dB) \\\hline
        BS-device and device-Rx channel & Rich Scattering \\\hline
        Noise power & $\sigma_{u,l}^2=\sigma_c^2=N_oB$\\\hline
    \end{tabular}
    
    \label{tab:simparam}
\end{table}

\begin{figure}[htb]
\centerline{\includegraphics[width=\columnwidth]{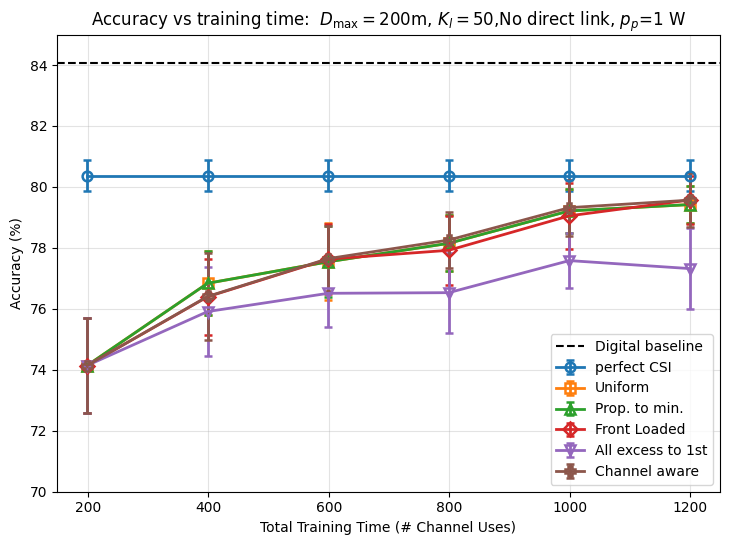}}
\caption{Accuracy vs. total training time for various training time allocation heuristics.  BS-Rx link blocked, $D_{max}=200$m, number of relay groups $L=3$, number of relay devices per group, $K_l=50$, pilot transmission power is $p_p=1$ W.}
\label{fig:AccvstauPk1Dmax200Pp1NDL}
\end{figure}

Fig. \ref{fig:AccvstauPk1Dmax200Pp1NDL} shows the classification accuracy vs. total training time ($\tau$) for relay power $P_k=1$ W, pilot power $p_p=1$ W, network size $D_{max}=200$m, and number of groups $L=3$. Direct BS-Rx link is assumed to be blocked.  In this plot, the performance for various training time allocation heuristics are observed. The results show that for high pilot power and $L=3$, most heuristics approach perfect-CSI performance given sufficient channel training time. Increasing training time significantly improves performance, but with diminishing returns. The best performing heuristics are Uniform, Proportional, and Channel-Aware heuristics. In this configuration, the Uniform and Proportional heuristics yield exactly the same allocation. The Channel-aware heuristic performs slightly better ($\sim 0.1\%$).   Over-allocating training time to a single hop leads to suboptimal performance and high variance due to imbalanced CSI quality across hops. Although the channel-aware allocation performs slightly better, the improvement is limited. This is because the channels are not very heterogeneous in this setting. 

\begin{figure}[thb!]
\centerline{\includegraphics[width=\columnwidth]{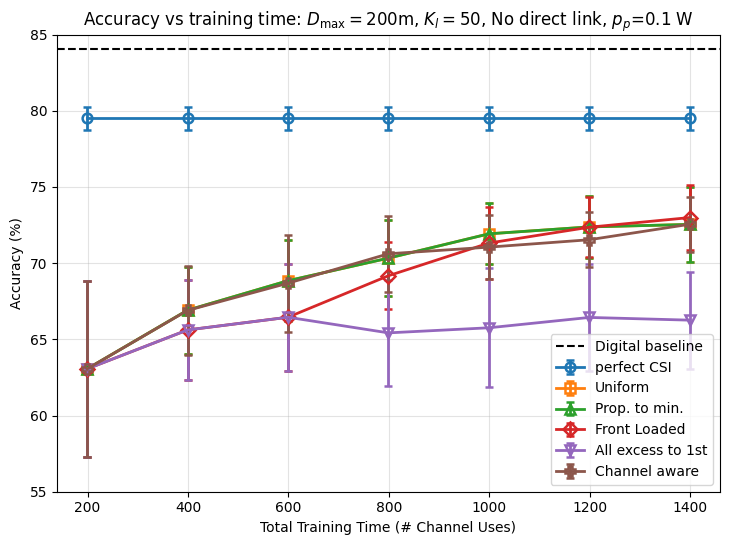}}
\caption{Accuracy vs. total training time for various training time allocation heuristics.  Bs-Rx link blocked, $D_{max}=200$m, number of relay groups $L=3$, number of relay devices per group, $K_l=50$, pilot transmission power is $p_p=0.1$ W. }
\label{fig:AccvstauPk1Dmax200Pp01NDL}
\end{figure}

Fig. \ref{fig:AccvstauPk1Dmax200Pp01NDL} shows the classification accuracy vs. total training time for a much smaller pilot power of $p_p=0.1$ W.  When the pilot transmission power is low, the performance is limited by channel estimation quality rather than training duration. While increasing the total training time improves accuracy, a significant performance gap with respect to perfect CSI remains even for large training budgets. In this regime, balanced training allocation strategies outperform highly asymmetric ones, and front-loaded allocation becomes more effective as channel training duration increases. In contrast, allocating all excess training to a single hop results in severe performance degradation due to uneven CSI quality across the multi-hop network.

\begin{figure}[t!]
\centerline{\includegraphics[width=\columnwidth]{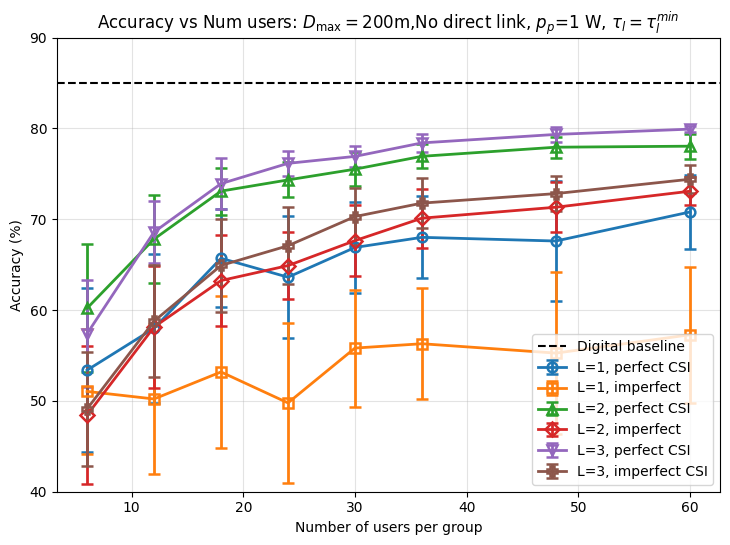}}
\caption{Accuracy vs. number of relay devices per group, for different number of groups.  Bs-Rx link is blocked, $D_{max}=200$m, AF relay power $P_k=1$ W, Pilot transmission power is $p_p=1$ W. Minimal training time is allocated for each hop. }
\label{fig:AccvsKPk1Dmax200Pp1NDL}
\end{figure}

Fig. \ref{fig:AccvsKPk1Dmax200Pp1NDL} shows the classification accuracy vs. the number of relay devices per group for different number of hops. Each hop is allocated just enough training time to obtain orthogonal pilots ($\tau_l=\tau_l^{min}$).  The results clearly show that near-perfect imitation accuracy can be achieved with a larger number of hops (e.g., $L=3$). Moreover, the variance is small and the performance is stable. However, for fewer hops (e.g., L=1), almost no improvement can be obtained by increasing the number of relays per group. Besides, the imitation accuracy exhibits a much higher variance.

\section{Conclusions}
In this paper, we investigated the impact of imperfect CSI on the performance of OTA implementation of a FC neural network layer over a multi-hop AF relay network. The considered architecture employs a multi-antenna transmitter and receiver, with relay devices organized into multiple groups to enable multi-hop transmission and mitigate pathloss over long distances. Channel estimation is performed using LS.

We proposed five heuristic strategies for allocating total channel training time across hops and analyzed their effect on the inference accuracy. Numerical results demonstrate that increasing the number of relay groups (e.g., $L=3$) helps shorten individual link distances and improves channel estimation quality. Under such configurations, with sufficient pilot power and training duration, the OTA implementation achieves accuracy levels close to that of the digital baseline.

Several directions remain for future work. First, synchronization errors among relay nodes may significantly impact system performance \cite{shao2021federated} and should be carefully analyzed and mitigated. Second, selecting an optimal subset of relays from a larger pool can improve system efficiency and scalability. Finally, incorporating energy harvesting capabilities at relay nodes and enabling simultaneous information transmission alongside OTA computation represent promising avenues for further research.

\bibliographystyle{IEEEtran}
\bibliography{OTAC}

\end{document}